\documentclass[12pt]{iopart}
\usepackage{iopams}
\usepackage[latin1]{inputenc}
\usepackage{graphicx}

\newcommand{\bfA}{{\bf A}}
\newcommand{\bfa}{{\bf a}}
\newcommand{\bfb}{{\bf b}}

\newcommand{\bfx}{{\bf x}}
\newcommand{\rmsol}{{\rm sol}}

\newcommand{\bsol}{b^{\rmsol}}
\newcommand{\bfxsol}{\bfx^{\rmsol}}
\newcommand{\bfbsol}{\bfb^{\rmsol}}
\newcommand{\ts}{{t_{\rm s}}}
\newcommand{\sn}{\sigma_{\rm noise}}
\begin{document}

\title[NNLC: Non-Negative Least Chi-square minimization]{NNLC: Non-Negative Least Chi-square minimization and application to HPGe detectors}

\author{ P Désesquelles$^{1}$, TMH. Ha$^{1}$, A Korichi$^{1}$, F Le Blanc$^{2}$ and CM Petrache$^{2}$}

\address{$^{1}$ CSNSM CNRS/IN2P3 and Université Paris Sud 11, 15 rue G. Clémenceau, 91405 Orsay, France}
\address{$^{2}$ IPNO CNRS/IN2P3 and Université Paris Sud 11, 15 rue G. Clémenceau, 91405 Orsay, France
\newline
on behalf of the AGATA collaboration}

\ead{Pierre.Desesquelles@in2p3.fr}

\begin{abstract}
A new method is proposed for the problem of solving chi-square minimization with a positive solution. This method is embodied in an evolution of the popular NNLS algorithm. Its efficiency with respect to residue minimization is illustrated by the improvement it permits on the location of gamma-interactions inside an AGATA HPGe detector.

\end{abstract}
\submitto{\JPG}
\maketitle

\section{Introduction}
Many experimental data analysis and model validation consist in solving a so-called inverse problem \cite{Tar}. Typically, one wishes to determine the distribution of a hidden variable knowing the distributions of measured variables and the transform function that connects both types of variables\footnote{For example, the determination of energy spectrum of the particles entering a detector from the shape of the delivered signals. In this case, the transform function is the response function of the detector.}. As a distribution is, by definition, a positive function, the algorithm used to solve this kind of inverse problem must converge towards the optimum positive solution. A very popular algorithm used to solve this problem is the Hanson and Lawson NNLS (Non-Negative Least Squares \cite{Law}\footnote{NNLS is implemented in the Matlab environment as lsqnonneg.}). However, this algorithm consists in minimizing the {\em residue} between the measured observable distribution and the theoretical distribution (i.e. the hidden distribution times the transform function). In most actual situations, the measured distributions and the transform matrices are affected by fluctuations and uncertainties. In such a case, what has to be minimized is not a residue but a {\em chi-square}. In the following, we show how these different kinds of uncertainties may be taken into account in the minimization. We introduce a new algorithm NNLC (Non-Negative Least Chi-square) based on NNLS which allows to handle uncertainties. In the second part, we apply both algorithms to a practical example. It will be shown how the new protocol improves the gamma hit location inside germanium detectors.

\section{Solving of the inverse problem}
\subsection{Formalism}

We consider a discreet inverse problem:

\begin{equation}
\label{Eq Ax=b}
\bfA\,\bfx = \bfb\,,
\end{equation}

\noindent where $\bfA$ is the transform matrix, $\bfx$ is the vector of unknowns and $\bfb$ is the (measured) vector of observations. When $\bfx$ represents the unknown distribution of a variable then each of its components has to be non-negative \cite{Des2}. On the other hand, the components of the transform matrix and of the observation matrix may be negative (this is the case if these distributions are detector signals). Most of the time, in real-life problems, no exact solution exists and what is searched is the non-negative solution $\bfxsol$ that minimizes the residue:
\begin{eqnarray}
\label{Eq R}
R^2 &=& \sum_{i} (b_{i}-\bsol_{i})^2 = \| \bfb-\bfA\,\bfxsol \|^2\ \, \\
\label{Eq bsol=Axsol}
 {\rm with}\ \, \bfbsol&=&\bfA\,\bfxsol\,.
\end{eqnarray}

\noindent In the following the $j^{\rm{th}}$ columns of matrix $\bfA$ will be considered as a vector and noted $\bfa_{j}$.

\subsection{The NNLS algorithm}

Such linear systems can be solved using the NNLS method. This fast iterative algorithm converges towards a positive $\bfxsol$ vector with a maximum of null components. Indeed, the initial guess for the solution is the null-vector. The main loop of the algorithm consists in adding one minimization component to the system solving until the stopping condition is reached. Thus, at the $k^{\rm{th}}$ iteration, the size of the matrix to invert is $k\times k$. At each iteration, the new component added to the system is the one that maximizes the gradient of the residue. The stopping condition is reached when the absolute value of this gradient is lower than a given tolerance. When the inversion gives negative components, an inner loop adds a kernel vector to the solution until it becomes non-negative.

\section{Chi-square minimization}
\subsection{Formalism}

The mere least square optimum solution of Eq.~(\ref{Eq Ax=b}) gives often an unsatisfactory solution since the components of the right hand side vector and of the transform matrix may be affected by fluctuations (noise, statistical fluctuations, \ldots), uncertainties or biases. In the most usual cases, what has to be minimized is not the residue, but a chi-square:

\begin{equation}
\label{Eq chi2}
\chi^2 = \sum_{i} \frac{\left( b_i - b^{\rmsol}_i\right)^2}{\sigma^2_{b_i} + \sigma^2_{b^{\rmsol}_i}}\ ,
\end{equation}

\noindent where $\sigma_{\bfb}$ and $\sigma_{\bfbsol}$ are the standard deviations of the fluctuations on $\bfb$ and $\bfbsol$. The first term is usually connected to the limitations of the detection system, the second term results from the uncertainties on the transform matrix, Eq. (\ref{Eq bsol=Axsol}).

We first consider the case when the transform matrix is not affected by uncertainties ($\sigma_{\bfbsol}=\bf 0$). Normalizing $\bfb$ and $\bfA$ by the standard deviation of the fluctuations of the observations: $b'_i = b_i/\sigma_{b_i}$ and $a'_{ij}= a_{ij}/\sigma_{b_i}$, one obtains $b'^{\rmsol}_i = \sum_{j} a'_{ij}\ x^{\rmsol}_{j}$. Hence, the chi-square can be written as a residue:

\begin{equation}
\label{Eq chi2=R2}
\chi^2 = \sum_{i} \left( b'_i - b'^{\rmsol}_i\right)^2 = \| \bfb'-\bfA'\,\bfxsol \|^2\ .
\end{equation}

Thus, least square minimization algorithms can also be used to solve this maximum likelihood problem. 

However, in many cases, the uncertainties (characterized by the standard deviations $ \sigma_{a_{ij}}$) on the transform matrix components cannot be neglected. These uncertainties may result from the fact that the transform matrix was obtained from a measurement or through a theoretical model that entails simplifications or hypotheses. Considering Eq. (\ref{Eq bsol=Axsol}), the uncertainties on the transform matrix induce a second denominator term in the chi-square:

\begin{equation}
\label{Eq sigma_bfbsol}
\sigma^2_{b^{\rmsol}_{i}} = \sum_{j} \sigma^2_{a_{ij}}\ {x_{j}^{\rmsol}}^{2}\,.
\end{equation}

Therefore, when the transform matrix is uncertain, the chi-square cannot be turned directly into a residue as in Eq. (\ref{Eq chi2=R2}), since $\sigma_{\bfbsol}$ depends on $\bfxsol$ which is the unknown to be determined. However, if the optimization algorithm used to solve the linear system works iteratively as it is most often the case (NNLS is iterative), the $\bfxsol$ obtained at iteration $k - 1$ can be used to calculate the fluctuations at iteration $k$. Hence, at each iteration we apply the following normalizations: 

\begin{eqnarray}
\label{Eq sigk}
\sigma_{i}^{(k)} &=& \sqrt{\sigma^2_{b_i}+\sum_{j} \sigma^2_{a_{ij}}\ {x^{(k-1)}_{j}}^2}\,,\\
\label{Eq bk}
b^{(k)}_i &=& \frac{b_i}{\sigma_{i}^{(k)}}\,, \\
\label{Eq ak}
a_{ij}^{(k)} &=& \frac{a_{ij}}{\sigma_{i}^{(k)}}\,,
\end{eqnarray}
so that:
\begin{eqnarray}
\chi^2 &=& \|\bfb^{(k)}-\bfA^{(k)}\,\bfx^{(k)}\|^2\,.
\end{eqnarray}

\noindent The chi square is finally transformed into a residue. 

\subsection{The NNLC algorithm}

In order to generalize the NNLS algorithm to inverse problems affected by uncertainties, we introduce the NNLC algorithm (Non-Negative Least Chi-square). As we have seen writing Eq (\ref{Eq chi2=R2}), when only the right hand side observation vector is affected by uncertainties, then the normalization of $\bfb$ and $\bfA$ can be made prior to the iterations. In the general case, the normalizations of Eqs. (\ref{Eq sigk}-\ref{Eq ak}) have to be included in the main loop. Some caution has to be taken when the uncertainties on $\bfA$ are large with respect to the uncertainties on $\bfb$. In this case, the normalization factor $\sigma_{i}^{(k)}$ may vary a lot from one iteration to the next one, Eq. (\ref{Eq sigk}). This means that the linear system changes at each iteration and that the minimization may not converge, or converge very slowly. When this effect appears, it can be solved by allowing a maximum variation of the standard-deviations of 10\% at each step.

In the following, we apply this new algorithm to the problem of locating the hits of gamma-rays inside a detector. 

\section{Application to the location of the interactions of a gamma-ray into a germanium crystal}

\subsection{Signal decomposition as an inverse problem}

A gamma-ray which enters a germanium detector may interact one or more times with the electrons of the crystal. In order to use this property for gamma detection, an electric field is applied to the crystal, so that each gamma-electron interaction (hit) provokes electron and hole cascades towards the anode and the cathode. The motions of the charges induce charge-signals at the electrodes. The shapes of the signals depend on the locations of the hits. When several hits occur simultaneously, the resulting signal is the sum of the signals induced by the individual hits. Moreover, the amplitude of the signal is proportional to the energy deposited by the gamma. Finding the locations and the energies of the hits allows to measure the energy and the direction of the gamma-ray \cite{Wae}. This is a widely used technique for modern gamma detection \cite{Ola,Cre,Dox,Kha1}.

The problem of determining the locations of the hits from the shape of the resulting signal \cite{Des3} is a typical inverse problem. The observable vector is the detected signal, the vector of unknowns contains the energy deposit on each point of the crystal and the transform matrix gives the shape of the signals for every point in the crystal. Of course, in order to keep finite size $\bfA$ and $\bfx$ matrices, the volume of the crystal is discretized into a finite number of voxels (in our case, we consider $2\times 2\times 2$~mm voxels for a crystal volume of about 400 cm$^3$). Each column $\bfa_{j}$ of the transform matrix is then the signal corresponding to a hit deposing a unit energy in the $j^{\rm{th}}$ voxel, and $x_{j}$ is the energy actually deposited. Due to the additive property of the signals, signal decomposition corresponds to the solving of Eq. (\ref{Eq Ax=b}).

\begin{figure}[htbp]
\begin{center}
\includegraphics[width=7cm]{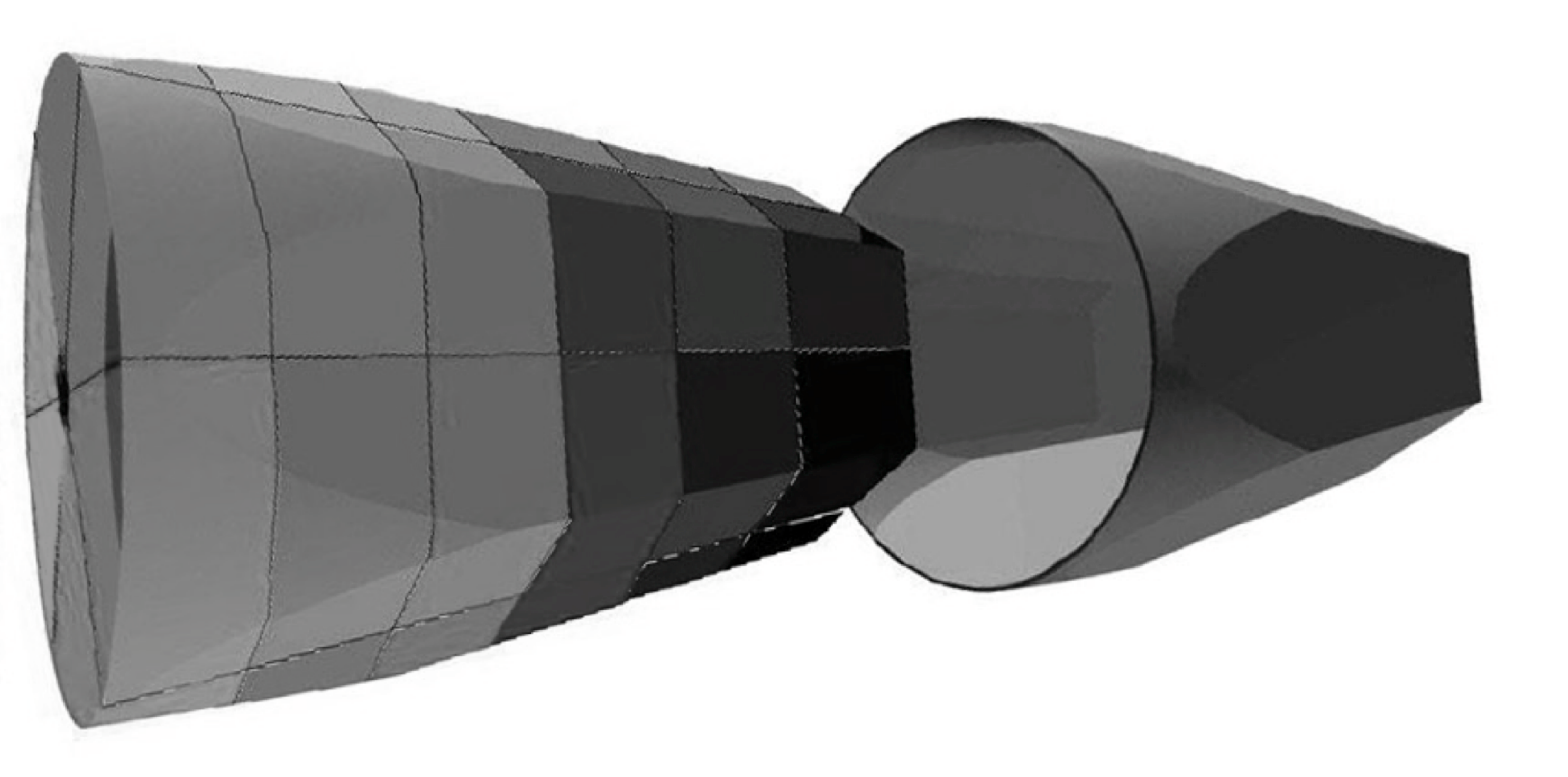}
\caption{AGATA germanium crystal and its capsule. The hole along the axis is the anode. The outer surface is covered by 36 cathodes.}
\label{Fig crystal}
\end{center}
\end{figure}

\begin{figure}[htbp]
\begin{center}
\includegraphics[width=7.8cm]{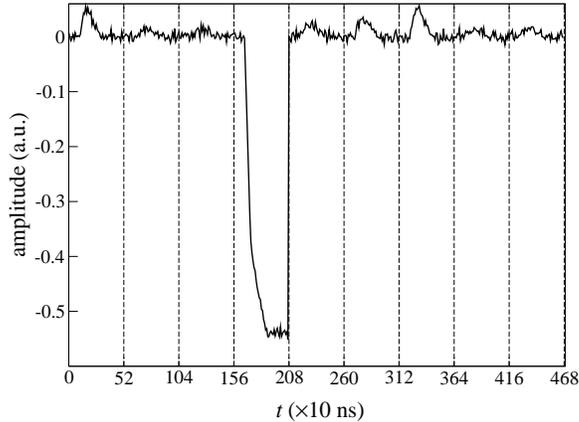}
\caption{Example of signal concatenation used to build the $\bfb$ and the $\bfa_{j}$ vectors. The 9 signals belong to the hit segment (fourth signal) and to its neighbors.}
\label{Fig metasignal}
\end{center}
\end{figure}

The AGATA \cite{Sim} germanium crystal used in this analysis is shown in Fig. \ref{Fig crystal}. The single anode is along its central axis. The outer surface is covered by six slices of six cathodes, thus the volume of the crystal is divided into 36 electrical segments. When a hit occurs in a given segment, the corresponding cathodes measures a net charge signal and the neighboring segments measure transient signals. All these signals, concatenated the one after the other, are used to form the $\bfb$ and the $\bfa_{j}$ vectors (see Fig. \ref{Fig metasignal}).

The $\bfa_{j}$ signals forming the transform matrix are obtain using the MGS simulation code \cite{Med} (they may also be obtained using a scanning device \cite{Kor,Bos}).

\subsection{Uncertainties}

\begin{figure}[htbp]
\begin{center}
\includegraphics[width=7.8cm]{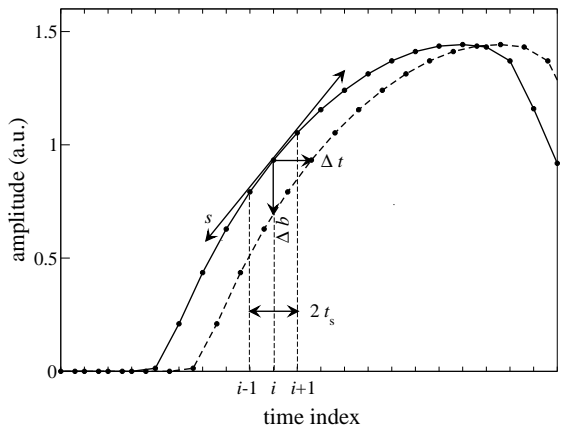}
\caption{The time shift $\Delta t$ induces an amplitude shift $\Delta b$ proportional to the slope $s$.}
\label{Fig Db_sDt}
\end{center}
\end{figure}

The main fluctuations on the detector signals are due to the electronic noise. This noise does not depend on the signal sample and corresponds to a standard-deviation $\sigma_{b_{i}} = \sn \approx 3$ keV. Moreover, the continuous signals delivered by the detector are time-discretized. As the discretizer is triggered when the noisy signal crosses a given threshold, the resulting discretized signals can be slightly time shifted \cite{Des4,Des6} (in our case, all the segment signals from a given event are translated by the same time shift). Typically, for $\ts = 10$~ns samples, the shift $\Delta t$ is of the order of some nanoseconds. The influence of the time shift on the amplitude in a given sample is illustrated in Fig. \ref{Fig Db_sDt}.

For small time jitters, the effect on the amplitude is proportional to the signal slope and to the time shift: $\Delta b_i =-s_i\,\Delta t$. The slope of the signal can be estimated by:

\begin{equation}
\label{Eq slope}
s_{i} = \frac{b_{i+1}-b_{i-1}}{2\,\ts}\ ,
\end{equation}

\noindent where $\ts$ is the sampling duration.

The slope calculated using the previous equation for the detected signals would be excessively affected by the noise (the fluctuations increase with the degree of derivation). Thus it is necessary to consider that the detected signal is the time reference, and that the transform matrix signals are time shifted with respect to the detected signal. This way, the slopes are calculated from smooth simulated signals. The fluctuations on the transform matrix elements are:

\begin{equation}
 \sigma_{a_{ij}} = \left|\frac{a_{i+1j}-a_{i-1j}}{2\,\ts}\right|\ \sigma_{\Delta t}\,,
\end{equation}
\noindent where $\sigma_{\Delta t}$ is the standard deviation of the time shifts \cite{Des4,Des6}.

Finally, from Eq. (\ref{Eq chi2}), the denominator of the chi-square reads:

\begin{equation}
\label{Eq denominator}
 \sigma^2_{\rm noise}+ \left(\frac{\sigma_{\Delta t}}{2\,\ts}\right)^2\ \sum_{j} \left(a_{i+1j}-a_{i-1j}\right)^2\ x_{j}^2\ .
\end{equation}

This expression of the standard deviation is declared to the NNLC algorithm.

\subsection{Results}

In order to compare the NNLS and NNLC algorithm results in this typical situation where both the observation vector and the transform matrix are altered by uncertainties, we simulate a large number of hits, at known random positions inside the crystal, for different values of the noise and of the time jitter standard deviation. We then apply both algorithms to calculate $\bfxsol$ and compare the precisions on the resulting hit locations.

\begin{figure}[htbp]
\begin{center}
\includegraphics[width=7.8cm]{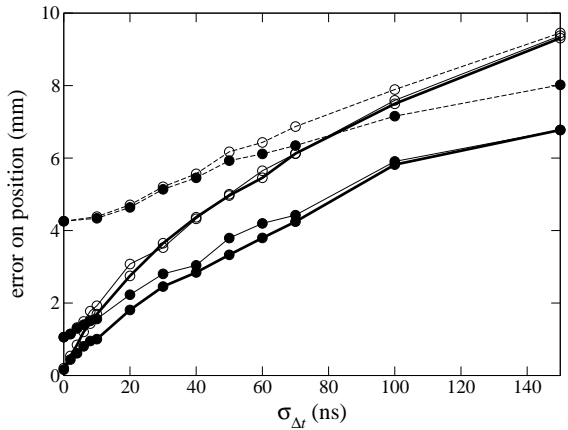}
\caption{Mean error on the location of the hits (averaged over 350 hits per point). The open dot curves are obtained using NNLS, the black dot curves are obtained using NNLC. The dotted lines correspond to 30 keV energy deposits, the thin lines to 300 keV deposits and the bold lines to 3 MeV deposits. The noise on the signals is 3 keV and the sample time is 10 ns.}
\label{Fig results}
\end{center}
\end{figure}

The precision of the location of the hit is presented in Fig. \ref{Fig results} as a function of the time jitter. The lower curves correspond to 3 MeV energy deposits, the middle curves to 300 keV deposits and the upper curves to 30 keV energy deposits. The error is lower for higher energies since the signal-to-noise ratio is larger. When there is no time jitter, the fluctuations are due to the noise, thus their standard deviation does not depend on the sample. In this case, the denominator can be factorized from the chi-square: $\chi^2 = (1/\sn^2) \sum_{i} (b_{i}-\bsol_{i})^2 \propto R^2$, see Eqs. (\ref{Eq chi2},\ref{Eq denominator}) and the chi-square and the residue minimizations give the same result. This is no longer true when the signals are affected by a time jitter even for small shifts (much lower than 10~ns, the sample duration). When the effect of the noise becomes negligible with respect to the effect of the time shift, the error on the position is almost independent on the energy deposit. For positive time jitters, the NNLC algorithm always gives better results than the NNLS algorithm and, as expected, the difference increases with the time shift.

\section{Conclusions}

This paper introduces a new method to account for the fluctuations and the uncertainties that may affect the transform function of an inverse problem as well as the measured observation vector. This method is embodied in the NNLC algorithm which is an evolution from the NNLS algorithm that replaces residue minimization by chi-square minimization. This new method was applied to the location of gamma-ray interactions inside germanium detectors. It was shown that, whatever the signal-to-noise ratio and the time jitter, the precision on the location is always better using NNLC. We believe that this very general method has a large number of actual applications.

\bibliographystyle{unsrt}

\end{document}